\documentclass[showpacs, twocolumn]{revtex4-1}
\usepackage{graphicx}
\usepackage{amsmath}

\usepackage{graphicx}  
\usepackage{dcolumn}   
\usepackage{bm}        
\usepackage{amssymb}   

\begin{document}

\title{Anderson localization of elementary excitations in disordered binary Bose mixtures: 
Effects of the Lee-Huang-Yang quantum and thermal corrections}

\author{Zohra Mehri$^{1}$ and Abdel\^{a}ali Boudjem\^{a}a$^{2}$}

\affiliation{$^1$ Department of Physics, Faculty of  Sciences and Technology, Ahmed Zabana University of Relizane, Bourmadia, P.O. Box 48000,Relizane, Algeria.\\
$^2$ Department of Physics, Faculty of Exact Sciences and Informatics, Hassiba Benbouali University of Chlef P.O. Box 151, 02000, Ouled Fares, Chlef, Algeria}

\email{a.boudjemaa@univ-chlef.dz}
\begin{abstract}
 
We investigate analytically and numerically the Anderson localization of quasiparticles in binary Bose mixtures in the presence of the Lee-Huang-Yang quantum and thermal corrections 
subjected to correlated disordered potentials. We calculate the density profiles, the Bogoliubov quasiparticles modes, and the localization length in both the mixture and droplet phases.
We show that for one-dimensional speckle potentials, the peculiar interplay of disorder and the Lee-Huang-Yang  fluctuations may
enhance localization in one component while inducing delocalization in the other. Our results reveal also that thermal fluctuations lead to the emergence 
of two and multiple localization maxima. 
In the droplet state,  our findings uncover that the Anderson localization of quasiparticles is weak in the flat-top region since 
its excitation modes are restricted to those below the particle-emission threshold. 

\end{abstract}

\maketitle

\section{Introduction}

Ultracold atoms in random environments have attracted  intense attention so far due to their unprecedented control over the disorder strength and interactions. 
One of the most intriguing phenomena in such disordered cold gases, is Anderson localization (AL) (see e.g. \cite {Lye, Billy, Roat, Laur,Yuk, Ski, Jen, Mc})
which is a direct consequence of interference between multiple-scattering paths \cite{Anderson}.     
The competition between disorder and interactions plays a crucial role in understanding many aspects of ultracold gases including: 
the Bose glass \cite{Ma, Giam, Fich, Scal, Krauth}, disordered  Bose-Einstein condensates (BECs) in optical lattices \cite {Schut,Wh, Pas, Deis},  
Bose-Fermi mixtures \cite {Ahuf,Franc}, and dipolar BEC in random potentials \cite{Krum, Nik,Ghab, Boudj8, Boudj9, Boudj10}. 
Such disordered systems open exciting new avenues for simulating a plethora of novel quantum phenomena.

Furthermore, disordered two-component BECs exhibit rich physics not accessible in a single-component BEC 
due to the peculiar interplay of intra- and interspecies interactions and disorder effects in both equilibrium and nonequilibrium states \cite{Nied,Xi,Mard,Boudj1,Boudj2,Boudj3}.
We have shown in addition that the disorder  plays a crucial role in controlling miscible-immiscible phase separation in binary Bose mixtures \cite{Boudj1}.
One of the major systems observed in Bose mixtures is the so-called self-bound droplets \cite{Petrov}. 
Their stabilization mechanism is due to a subtle balance of attractive mean-field interactions and repulsive forces provided by the Lee-Huang-Yang (LHY) quantum fluctuations \cite{LHY}. 
Most recent works have demonstrated that the disorder may split the droplet into several mini droplets and eventually dissolve it \cite{Sahu, Boudj4,Boudj5, Boudj6}.

On the other hand, AL of Bogoliubov quasiparticles (BQPs) in a weakly interacting Bose gas with weak random potentials has been studied in 
one (1D)- and higher dimensions using different approaches \cite{Pav, Laurent1,Laurent2, Gaul,Lell}. 
These studies revealed that the localization length reduces to the one of phonons at low energy and to that of non-interacting particles at high energy \cite{Pav, Laurent2}.
However, AL of BQPs in binary BECs remains largely unexplored. 
A detailed description of these BQPs have been given in \cite{FedG, Boudj7} using different methods, which takes into account  pair excitations, 
and thus, generalizes the standard Bogoliubov theory. 
The excitation spectrum of binary BECs is composed of two elementary branches namely: 
density excitations corresponding to the highest energy branch and the spin modes corresponding to the lowest energy branch. 
Elementary excitations constitute an important source of information for understanding the physics of Bose mixtures such as 
normal and anomalous quantum fluctuations, correlation functions, superfluidity and thermodynamic properties.
At finite temperatures where a major fraction of the atoms are no longer in the condensate state, such elementary excitations can play a crucial role. 
It is therefore instructive to look at how the intriguing interplay between the disorder and such elementary excitations affects the localization properties of Bose mixtures and quantum droplets. 
Moreover, competition/cooperation of the LHY quantum and thermal effects with disorder in Bose mixtures is still an unanswered question. 

The aim of this paper is to theoretically investigate the AL of BQPs in a weakly interacting Bose mixture with the LHY quantum and thermal corrections subjected to 
weak and correlated random potentials. 
In our paper we focus on a 1D geometry since AL is proved to be robust in lower dimensions.
It has been shown that the inclusion of these LHY corrections leads to reduce the disorder effects and hence modify the localization process in a disordered single BEC \cite{Boudj11,Nagler}. 
From a theoretical point of view, addressing such a question is not an easy task due to the competition between intra- and interspecies interactions, quantum fluctuations and the disorder.
Emphasis is set here on the determination of the BQP modes, the localization length, and the Lyapunov exponent associated with each component. 
To this end, we use the density-phase formalism in the hydrodynamical approach based on the finite-temperature generalized Gross-Pitaevskii equation  (FTGGPE)
which includes extra nonlinearities stemming  from the LHY quantum and thermal fluctuations. 
We calculate the density profiles of each species by numerically solving the coupled FTGGPE.  

It is shown that if the disorder potential constitutes a weak perturbation, the Bogoliubov-de-Gennes equations (BdGE) governing the transport of the BQPs transform into
a Schr\"odinger-like equation that describes the scattering properties similarly to the noninteracting particles \cite{Laurent1,Laurent2}.
Nevertheless,  AL of BQPs discussed here differ qualitatively from the noninteracting case due to their phonon character at low momenta in both components and even in the droplet state. 
The solution of the obtained effective Schr\"odinger equation yields a screening term which encodes interactions and the LHY corrections.
It is found that these latter tend to alter the screening functions in both the phonon and single-particle regimes.

For a 1D speckle potential, our results reveal that the LHY quantum and thermal fluctuations may affect the localization picture.
Strong enough thermal fluctuations induce several localization maxima in the lower excitations component and eventually prevent the localization.
Finally, we discuss the existence of the AL of BQPs in the droplet regime under the influence of speckle potentials.
We show that BQPs are less localized in the flat-top (plateau) region. 

The remainder of the paper is structured as follows.
In Sec.~\ref{Mod},  we introduce the coupled FTGGPE model and discuss its pertinence for disordered Bose mixtures.
Section \ref{PerTh} is devoted to the perturbation theory which is based on the coupled FTGGPE. We calculate the density profiles of each component in
different regimes.
In Sec. \ref{SolBdG}  we present a general solution of the BdGE which transforms these latter into an effective Schr\"odinger-like equation describing the scattering
and localization properties of the BQPs.
We address the effects of quantum and thermal fluctuations on the excitations spectra and on the Bogoliubov modes.
Section \ref{ALE} aims to investigate AL of BQPs in weakly interacting Bose mixtures subjected to 1D speckle potentials using the above method.
We compute in particular the localization length and the screening term which encodes interactions and the LHY quantum and thermal corrections. 
In  Sec.~\ref{ALQL} we extend our study to disordered quantum droplets and verify the existence of AL of BQPs.
Finally, Sec.~\ref{Conc} concludes the paper.

\section{ Model} \label{Mod}

We consider a weakly interacting 1D two-component BEC  with equal atomic masses $m_1=m_2=m$  confined in an external random potential $U(z)$ at finite temperatures.
We assume that the condensates vary slowly at the scale of the extended healing length \cite{Boudj4} and 
the disorder potential changes smoothly in space on a length scale comparable to the healing length \cite{Boudj4,Boudj11,Nagler}.
Therefore, the dynamics of such a system is governed by the following coupled FTGGPE \cite{Boudj5,Boudj11},  which is a classical-field equation:
\begin{align} \label{GPE}   
i\hbar \frac{\partial \phi_j}{\partial t}&=\bigg[-\frac{\hbar^2}{2m} \frac{\partial^2}{\partial z^2} + U -\mu_j +g_j |\phi_j|^2+ g_{12} |\phi_{\overline j}|^2 \\
&+\alpha_j\, |\phi_j|+\alpha_j^T |\phi_j|^{-3}\bigg] \phi_j, \nonumber
\end{align}
where $\overline j=3-j$, $\phi_j=\phi_j(z,t)$ is the wavefunction of each condensate,  and $\mu_j$ is the chemical potentials related to bosonic components.
The coefficients $g_j=(4\pi \hbar^2/m) a_j$ and $g_{12}=g_{21}= (4\pi \hbar^2/m) a_{12}$ with 
$a_j$ and $a_{12}$ being the intraspecies and the interspecies scattering lengths, respectively. 
The last two terms in Eq.~(\ref{GPE}) account for the LHY quantum and thermal corrections \cite{LHY} stemming from the normal and anomalous (pairing) fluctuations \cite{Boudj5}.
The quantum, $\alpha_j$, and thermal, $\alpha_j^T$, LHY strengths are determined by the excitation spectrum, and hence depend on intra- and interspecies interactions.
Note that our coupled FTGGPE (\ref{GPE}) hold for a system of general dimension.

It is convenient to introduce the density-phase formalism which requires to write the condensate wavefunction  \cite{Lell,Boudj111} as: $\phi_j (z,t) = \sqrt{n_{cj}(z,t)} e^{i \theta_j(z,t)}$, where 
$n_{cj}(z,t)=|\phi_j(z,t)|^2$ is the gas density of each condensate, and $\theta_j (z,t)$ encodes the phase which is real and related to the superfluid velocity as $v_j=(\hbar/m) \nabla \theta_j$. 
Substituting this transformation into the coupled FTGGPE (\ref{GPE})  and separating real and imaginary parts, one obtains 
the continuity and the Euler-like equations, respectively
\begin{equation}  \label{Hyd1} 
\frac{\partial n_{cj}} {\partial t} +\frac{\partial }{\partial z}(n_{cj} v_j)=0, 
\end{equation}  
and 
\begin{align}  \label{Hyd2} 
m\frac{\partial v_j}{\partial t} &=-\frac{\partial }{\partial z} \bigg [-\frac{\hbar^2}{2m \sqrt{n_{cj}} } \frac{\partial^2 \sqrt{n_{cj}}}{\partial^2 z} +\frac{1}{2} m v_j^2 +U-\mu_j \nonumber\\
&+ g_j n_{cj} +g_{12} n_{c\overline j} +\alpha_j\, n_{cj}^{1/2}+\alpha_j^T n_{cj}^{-3/2}\bigg], 
\end{align}
where $\left(\partial^2 \sqrt{n_{cj}}/ \partial z^2\right)/ \sqrt{n_{cj}}$ represents the so-called quantum pressures.

Assuming small fluctuations of the density $ n_{cj} (z,t)=n_j (z)+\delta n_{cj} (z,t)$, where $\delta n_{cj} /n_j\ll 1$, 
we expand Eqs.~(\ref{Hyd1}) and (\ref{Hyd2})  with respect to $\delta n_{cj} $, and  $\partial \theta_j/\partial z$ around the stationary solution.\\
To zeroth-order, we obtain the static FTGGPE:
\begin{align}  \label{GPE0}
&\bigg[-\frac{\hbar^2}{2m} \frac{\partial^2}{\partial z^2}  + U(z) -\mu_j +g_j n_j(z)+ g_{12} n_{\overline j} (z)\\
&+\alpha_j\, n_j^{1/2} (z)+\alpha_j^T n_j^{-3/2} (z)\bigg]\sqrt{n_j (z)}=0. \nonumber
\end{align}
The solution of these coupled equations gives the disordered density background in each component, $n_j$.\\
First-order terms provide the following equations:
\begin{widetext}
\begin{align}
\hbar \frac{\partial} {\partial t} \frac{\delta n_{cj} (z,t)} {\sqrt{n_j(z)}} =&\bigg[-\frac{\hbar^2}{2m} \frac{\partial^2}{\partial z^2}  + U(z) -\mu_j + g_j n_j (z) + g_{12} n_{\overline j}(z)    
+\alpha_j\, n_j^{1/2} (z)+\alpha_j^T n_j^{-3/2} (z) \bigg] 2\sqrt{n_j (z)} \, \theta_j (z,t),  \label{F:td1}\\
2\sqrt{n_j (z)}\,\hbar \frac{\partial \theta_j (z,t)} {\partial t} &=-\bigg[ -\frac{\hbar^2}{2m} \frac{\partial^2}{\partial z^2}  + U (z)-\mu_j +3 g_j n_j (z) 
+ g_{12} n_{\overline j} (z) +2\alpha_j\, n_j^{1/2} (z)-2\alpha_j^T n_j^{-3/2} (z)\bigg] \frac{\delta n_{cj} (z,t)} {\sqrt{n_j (z)}}  \nonumber\\
&- 2 g_{12}  \sqrt{n_j (z) n_{\overline j} (z) } \, \frac{\delta  n_{c\overline j} (z,t)}{\sqrt{n_j (z)}}.    \label{F:td2} 
\end{align}
\end{widetext}

Expanding the density and the phase in the basis of the Bogoliubov excitations:
\begin{subequations}\label{B:td0}
\begin{align}
&\theta_j(z,t)=\frac{-i}{2\sqrt{n_j(z)}}  \sum\limits_k [f_{jk}^{+} (z) e^{-i \varepsilon_k  t/ \hbar } \hat {b}_{jk}-\text {H.c.}], \label{B:td01}\\
&\delta  n_{cj}(z,t)=\sqrt{n_j(z)} \sum\limits_k [f_{jk}^{-}(z) e^{-i \varepsilon_k  t/\hbar } \hat {b}_{jk}+\text {H.c.}], \label{B:td02}
\end{align}
\end{subequations}
where $\hat {b}_{k}$ is the annihilation operator of an excitation of energy $\varepsilon_k$,
one then finds that wavefunctions $f_{jk}^{\pm} (z)$ obey the BdGE:
\begin{subequations}\label{B:td}
\begin{align}  
\varepsilon_k f_{jk}^{-} (z)&=\bigg [-\frac{\hbar^2}{2m} \frac{\partial^2}{\partial z^2}  + U(z) -\mu_j+g_j n_j(z) \label{B1:td}  \\
&+ g_{12} n_{\overline j} (z) +\alpha_j\, n_j^{1/2}(z)+\alpha_j^T n_j^{-3/2}(z)\bigg] f_{jk}^{+}(z),   \nonumber\\ 
\varepsilon_k f_{jk}^{+}(z)&=\bigg [-\frac{\hbar^2}{2m} \frac{\partial^2}{\partial z^2}  + U(z) -\mu_j +3g_j n_j(z)\label{B2:td} \\ 
&+g_{12} n_{\overline j} (z)+2\alpha_j\, n_j^{1/2} (z) -2\alpha_j^T n_j^{-3/2} (z)\bigg] f_{jk}^{-}(z)  \nonumber\\
&+ 2 g_{12} \sqrt{n_j(z) n_{\overline j}(z)} f_{\overline j k}^{-}(z), \nonumber
\end{align}
\end{subequations}
with the normalization $\int d z [f_{jk}^{+} {f_{jk'}^{-}}^*+ f_{jk}^{-} {f_{jk'}^{+}}^*]=2\delta_{k,k'}$.
Equations (\ref{GPE0}) and (\ref{B:td}) form a complete set to calculate the ground-state and collective modes of  binary  BECs.

\section{Perturbation theory} \label{PerTh}

Assuming that $U$ is a weak external potential with a vanishing ensemble average $\langle U(z)\rangle=0$ 
and a finite autocorrelation of the form $R (z-z')=\langle U(z) U(z')\rangle$. 
In the dimensionless form $R (z)=U_R^2 c_2(z/\sigma)$, where $U_R$ and $\sigma$ are respectively, the amplitude and the correlation length of $U$. 
If the extended healing length, $\xi_j$ is much smaller than the size of the system $L$, the density remains delocalized for a weak disorder \cite{Laurent1, Laurent2}. 
It is then convenient to solve Eq.~(\ref{GPE0}) perturbatively.

For a given chemical potential we suggest the expanssion:
\begin{equation} \label{pertur}
\sqrt{n_j(z)}=\sqrt{n_{j}}\bigg[\phi_j^{(0)} +\phi_j^{(1)}(z)+\phi_j^{(2)}(z)+ \cdots\bigg],
\end{equation}
where 
$$\sqrt{n_{j}}= \sqrt{ \left(\mu_j-g_{12} \phi^{(0)2}_{\overline j} -\alpha_j |\phi_j^{(0)}|- \alpha_j^T |\phi_j^{(0)}|^{-3}\right)/g_j }$$
is the uniform solution in the absence of a disorder potential, and $\phi^{(i)}$ signals the $i^{th}$-order contribution with respect to the disorder potential.
They can be determined by inserting the perturbation expansions (\ref{pertur}) into the FTGGPE (\ref{GPE0}).  Gathering terms up to $U^2$, we find
	\begin{equation} \label{n0}
		\phi_j^{(0)}=1,
	\end{equation}
		\begin{equation} \label{n1}
\phi_j^{(1)}=\bigg(-\frac{1}{2\mu_{0j}}(G_{\xi_j}*U)+\frac{1}{2\mu_{0\overline{j}}}G_{\xi_j}*(G_{\xi_{\overline{j}}}*U)\frac{g_{12}n_{\overline{j}}}{\mu_{0j}}\bigg)A_j,
	\end{equation}
and 
		\begin{align} \label{n2}
	\phi_j^{(2)}&=-\frac{1}{2}G_{\xi_j}*\bigg[\frac{U} { \mu_{0j}}\phi_j^{(1)}+\frac{3g_jn_{j}}{\mu_{0j}}\phi_j^{(1)}\phi_j^{(1)} \\
&+\frac{g_{12}n_{\overline{j}}}{ \mu_{0j}}\bigg(\phi_{\overline{j}}^{(1)}\phi_{\overline{j}}^{(1)}+2\phi_{\overline{j}}^{(1)}\phi_{j}^{(1)}\bigg) \nonumber\\
&+\frac{\alpha_j n_{j}^{1/2}}{\mu_{0j}}\phi_{j}^{(1)}\phi_{j}^{(1)}+\frac{\alpha_j^T n_{j}^{-3/2}}{4\mu_{0j}\phi_j^{(1)}\phi_j^{(1)}}\bigg]-G_{\xi_j}*\frac{g_{12}n_{\overline{j}}}{\mu_{0j}}\phi_{\overline{j}}^{(2)}\nonumber,
	\end{align}
where $\mu_{0j}=g_j n_{j}(1+\nu_j/2-3\nu_j^T/2)$,  $\nu_j= \alpha_j n_{j}^{1/2}/g_jn_{j}$, $\nu_j^T=\alpha_j^T n_{j}^{-3/2}/g_jn_{j}$, 
$ G_{\xi_j}$ is the Green function associated with the operator, $-\xi_j (\partial^2/\partial z^2)+1$, in Fourier space takes the form:
$G_{\xi_j}(k)=(1+\xi_j^2k^2)^{-1}$, $\xi_j=\hbar/\sqrt{4m\mu_{0j}}$,
and
$$A_j=\frac{\Delta(1+\nu_j/2-3\nu_j^T/2)(1+\nu_{\overline{j}}/2-3\nu_{\overline{j}}^T/2)}{\Delta(1+\nu_j/2-3\nu_j^T/2)(1+\nu_{\overline{j}}/2-3\nu_{\overline{j}}^T/2)
-G_{\xi_j}*G_{\xi_{\overline{j}}}},$$ 
with $\Delta=g_j g_{\overline{j}}/g_{12}^2$  being the standard miscibility parameter.     

Consequently, the mean-field density term for an assymmetric mixture reads
	\begin{equation} \label{dens}
	n_j(z)=n_{j}\bigg[1+\bigg(-\frac{\tilde{U}_j}{\mu_{0j}}+\frac{ g_{12} n_{ \overline j}} { \mu_{0j} \mu_{0 \overline{j}} } G_{\xi_j}*\tilde{U}_{\overline{j}}\bigg)A_j\bigg],
	\end{equation}
where 
	\begin{align}
		\tilde{U}_j(z)&=\frac{ \mu_{0j} } {n_{j}} \left(\langle n_j \rangle -n_j(z) \right) \\
&+g_{12} n_{\overline{j}}G_{\xi_j} \bigg(\frac{ \langle n_{\overline{j}} \rangle-n_{\overline{j}}(z)}{n_{{\overline{j}}}}\bigg),  \nonumber
	\end{align}
describes the modulations of the density due to the disorder with vanishing average  $\langle \bar{U} (z) \rangle=0$.
Note that $\tilde{U}(z)$ vanishes for $U=0$, and remain small for a weak external potential.
For $g_{12}=\alpha_j=\alpha_j^T=0$, the density reduces to that found  for a disordered single BEC \cite{Laurent2}.

Importantly, one can see from Eq.~(\ref{dens}) that the presence of disorder, quantum and thermal fluctuations may strongly affect the miscibility condition for binary Bose mixtures
\cite{Boudj1,Boudj7, Ota}. This means that the mixture undergoes a transition to an immiscible phase even if the mixture is initially miscible
and even in the absence of disorder potentials and vice versa (see below). 

\begin{figure}
\center
\includegraphics  [scale=0.29] {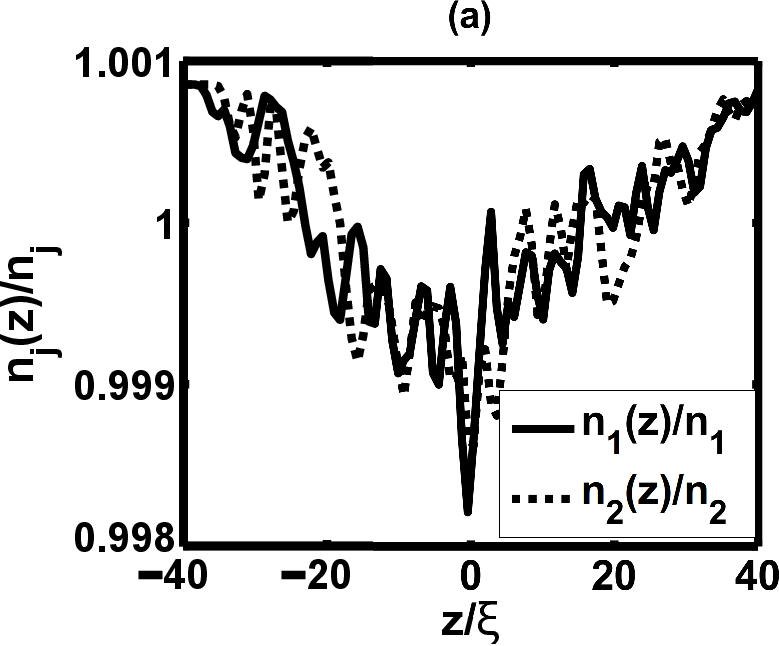}
\includegraphics  [scale=0.29] {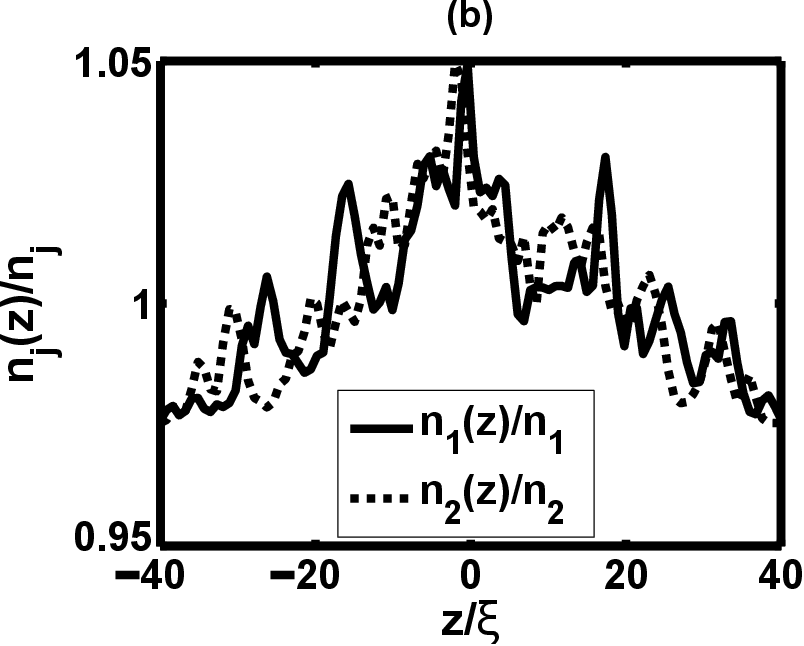} 
\includegraphics  [scale=0.29] {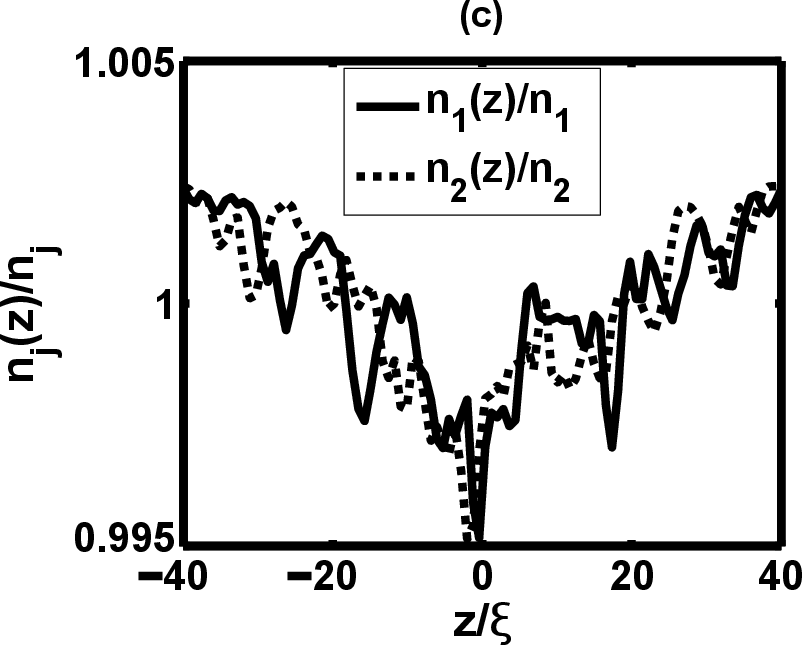} 
\includegraphics  [scale=0.29] {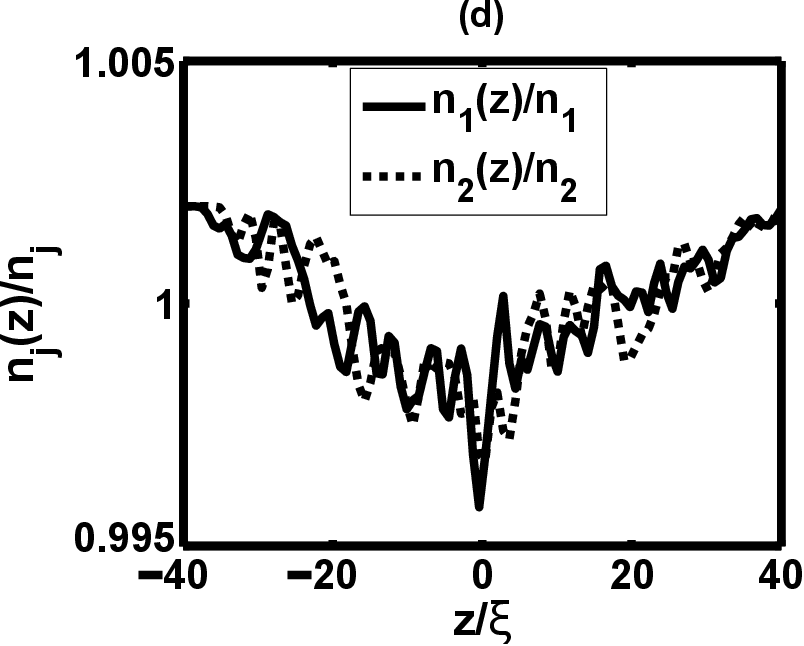} 
\caption{ Density profiles in the case of a 1D speckle potential with reduced autocorrelation function $c(z/\xi) = \sin(z/\xi)^2/ (z/\xi)^2$, 
obtained from the numerical solutions of Eq.~(\ref{dens}) for $\sigma/\xi=1$, $U_R/n_j g_j=0.1$, $g_{12}/g_1=0.5$, and $g_{12}/g_2=0.55$.
Parameters are: (a)  $\Delta=1.5$,  $\nu_1=\nu_2=0.2$, and $\nu_1^T=\nu_2^T=0.2$. 
(b) $\Delta=0.6$,  $\nu_1=\nu_2=0.2$, and $\nu_1^T=\nu_2^T=0.2$.
(c) $\Delta=1.5$,  $\nu_1=\nu_2=0.8$, and $\nu_1^T=\nu_2^T=0.2$.
(d) $\Delta=1.5$,  $\nu_1=\nu_2=0.2$, and $\nu_1^T=\nu_2^T=0.8$.}
\label{Dprof}
\end{figure}

Examples of the stationary solution of Eq.~(\ref{dens}) for both components, $j=1,2$, in the case of a 1D speckle potential with reduced autocorrelation function 
$c(z/\xi) = \sin(z/\xi)^2/ (z/\xi)^2$, and for arbitrary values of $\nu$ and $\nu^T$ are displayed in Fig.~\ref{Dprof}. 
We see that when the strengths of quantum and thermal fluctuations are equal ( i.e. $\nu_1=\nu_2$ and $\nu_1^T=\nu_2^T$), 
the two densities are almost comparable and thus, the mixture exhibits a broad area where the two components coexist even for $\Delta <1$.
The example in Fig.~\ref{Dprof} shows also that for $\Delta >1$, the densities develop a deep dip near the center, which is originated due to a peak of large amplitude in the disordered potential.
However, the densities exhibit a hump structure close to the center in the limit $\Delta <1$ where the disorder is dominated by the LHY corrections induced by interactions.
One can infer that the interplay of disorder,  intra-, interspecies interactions and the LHY quantum and thermal fluctuations may crucially affect 
the density profiles and the miscibility of the mixture.

Note that the perturbation expansions, $\phi_j^{(0)}$, $\phi_j^{(1)}$, and $\phi_j^{(2)}$ allow us also to calculate the disorder contribution to the equation of state (EoS)
and to the compressibility in a straightforward manner (see Appendix A).

\section{ Solutions of the  B\lowercase{d}GE and effective Schr\"odinger equation} \label{SolBdG}

In this section  we will be looking at solutions of the  BdGE  and derive the effective Schr\"odinger equation describing the scattering
and localization properties of the BQPs in disordered binary Bose mixtures.

Inserting the ground-state density (\ref{dens}) into Eqs.~(\ref{B:td}),  the coupled BdGE may be put into matrix form:
\begin{widetext}
\begin{align} \label{BdGE2}
		\frac{\hbar^2}{2m} \frac{\partial ^2}{\partial z^2}\begin{pmatrix} 
		f_{1k}^+\\
		f_{1k}^- \\
		f_{2k}^+\\
		f_{2k}^- \\
		\end{pmatrix}
&=\begin{pmatrix}
		0 & -\varepsilon_k &0&0\\
		-\varepsilon_k &2\mu_{01}&0&2g_{12}\sqrt{n_{1}n_{2}}\\
		0 & 0&0&-\varepsilon_k \\
		0&2g_{12}\sqrt{n_{1}n_{2}}&	-\varepsilon_k &2\mu_{02}\\
		\end{pmatrix}
		\begin{pmatrix}
		f_{1k}^+\\
		f_{1k}^- \\
		f_{2k}^+\\
		f_{2k}^- \\
		\end{pmatrix} \\
&+\begin{pmatrix}
		(U+B_1\tilde{U}_{1}+C_1\tilde{U}_{2})& 0&0&0\\
		0 &( U+D_1\tilde{U}_{1}+M_1\tilde{U}_{2})&0&Q_1 \\
		0& 0&	(U+B_2\tilde{U}_{2}+C_2\tilde{U}_{1})&0\\
		0 &Q_2&0&( U+D_2\tilde{U}_{2}+M_2\tilde{U}_{1})\\
		\end{pmatrix}
		\begin{pmatrix}
		f_{1k}^+\\
		f_{1k}^- \\
		f_{2k}^+\\
		f_{2k}^- \\
		\end{pmatrix}, \nonumber
\end{align}
	where\\
	$B_j=\bigg(\frac{g_{12}^2 n_{j}n_{\overline{j}}}{\mu_{0j} \mu_{0\overline {j}}}G_{\xi_{\overline{j}}}-1\bigg)A_j$, \\
	$C_j=\frac{g_{12}n_{\overline{j}}}{\mu_{0\overline{j}}}(G_{\xi_j}-1)A_j$,\\
$D_j=\bigg(3-\frac{g_{12}^2n_{j}n_{\overline{j}}}{ \mu_{0j} \mu_{0\overline{j}}}G_{\xi_{\overline{j}}}-\frac{ \alpha_j n_{j}^{1/2}} {2\mu_{0j}}
+\frac{15\alpha_j^T n_{j}^{-3/2}}{2\mu_{0j}}\bigg) A_j$, \\
	$M_j=\bigg[\bigg(3-\frac{\alpha_j n_{j}^{1/2}}{2\mu_{0j}}+\frac{15\alpha_j^T n_{j}^{-3/2}}{2\mu_{0j}} \bigg)G_{\xi_{j}}-1\bigg] A_j$, 
and
	$$Q_j=g_{12}\sqrt{n_{j}n_{\overline{j}}}\bigg[\tilde{U}_j \bigg(\frac{-1}{\mu_{0j}}+\frac{g_{12}n_{j}}{\mu_{0j} \mu_{0\overline{j}}}G_{\xi_{j}}\bigg) A_j
        +\tilde{U}_{\overline{j}}\bigg(\frac{-1}{\mu_{0 \overline{j}}}+\frac{g_{12}n_{\bar{j}}}{\mu_{0j} \mu_{0\overline{j}}}G_{\xi_{\overline{j}}}\bigg)A_j
+\frac{1}{2}\bigg(\frac{-\tilde{U}_j}{\mu_{0j}}+\frac{g_{12}n_{\overline{j}}}{\mu_{0j} \mu_{0 \overline{j}}}G_{\xi_{j}}*\tilde{U}_{\overline{j}}\bigg)\bigg(\frac{-\tilde{U}_{\overline{j}}}{\mu_{0\overline{j}}}+\frac{g_{12}n_{\overline j}}{\mu_{0j} \mu_{0\overline{j}}}G_{\xi_{\overline{j}}}*\tilde{U}_j\bigg) A_j^2\bigg].$$
To solve the BdGE (\ref{BdGE2}) we suppose that the density and the Bogoliubov functions $f_j^+$ and  $f_j^-$ tend to zero at the system boundaries.

For a homogeneous mixture ($ U(z)=0$ and $\xi_{\pm}/L \rightarrow 0$), the second term in Eq.~(\ref{BdGE2}) vanishes, 
thus the Bogoliubov excitations spectrum of a Bose mixture with the LHY quantum and thermal corrections takes the form:
\begin{equation} \label {Bog}
\varepsilon_{k\pm }= \sqrt{E_k^2+2E_k \mu_{\pm} }, 
\end{equation}
where $E_k= \hbar^2k^2/2m$, and
\begin{align}
	\mu_{\pm}&=\frac{g_1n_1+\frac{1}{2} \alpha_1 n_1^{1/2}-\frac{3}{2}\alpha_1^T n_{1}^{-3/2}}{2}\Bigg[1+\frac{g_2n_2+\frac{1}{2} \alpha_2 n_2^{1/2}-\frac{3}{2}\alpha_2^T n_{2}^{-3/2}}{g_1n_1+\frac{1}{2}\alpha_1 n_1^{1/2}-\frac{3}{2}\alpha_1^T n_{1}^{-3/2}}\\
	&\pm\sqrt{\bigg(1-\frac{g_2n_2+\frac{1}{2} \alpha_2 n_2^{1/2}-\frac{3}{2}\alpha_2^T n_{2}^{-3/2}}{g_1n_1+\frac{1}{2} \alpha_1 n_1^{1/2}-\frac{3}{2}\alpha_1^T n_{1}^{-3/2}}\bigg)^2
	+4 \Delta^{-1}\frac{g_2n_2}{g_1n_1 \big(1+\alpha_1 n_1^{1/2}/(2g_1n_1)-3\alpha_1^T n_{1}^{-3/2}/(2g_1n_1)\big)^2} }\Bigg].\nonumber
	\end{align}
\end{widetext}
The presence of the interspecies interactions $g_{12}$ gives rise to hybridize the two Bogoliubov modes namely : 
the upper branch, $\varepsilon_{k+}$, corresponds to the density modes  and the lower energy branch, $\varepsilon_{k-}$, is of spin nature. 
For $\alpha_j=\alpha_j^T=0$, the two excitation branches (\ref{Bog}) take the standard form of the Bogoliubov spectrum of two-component BECs.
For  $g_{12}= \alpha_j=\alpha_j^T=0$, $g_1 = g_2$ and $n_1 = n_2$, the dispersion relation (\ref{Bog}) reduces to that of the usual single BEC.
The two branches present a free-particle behavior at large $k$ and a phonon dispersion at low $k$,
$\varepsilon_{k \pm}=\hbar c_{s\pm} k$, where the sound velocities $c_{s\pm}$ are defined as:
\begin{equation}\label{CH2:Eq46}
c_{s\pm}= \sqrt{ \mu_{\pm}/m}.
\end{equation}

Solving the BdGE (\ref{BdGE2}) in the basis ($f_{1k}^{\pm}, f_{2k}^{\pm}$) is difficult due to their strong coupling arising from the weak disorder and interspecies interactions.
To circumvent this hindrance, we introduce the following linear transformation \cite{Laurent1, Laurent2}: 
\begin{equation}
g_{jk}^{\pm}=\pm{\rho}_{jk}^{\pm 1/2}f_{jk}^++{\rho}_{jk}^{\mp1/2}f_{jk}^-,
\end{equation}
where
	\begin{equation}
{\rho}_{jk}^{\pm}=\frac{\mu_{0j} }{\varepsilon_{k\pm}}+\sqrt{1+\bigg(\frac{\mu_{0j}} {\varepsilon_{k\pm}}\bigg)^2}.
	\end{equation}
In the basis of the $g_{j k}^{\pm}$ functions, the BdGEs take the form:
\begin{widetext}
\begin{align}
\frac{\hbar^2}{2 m} \frac{\partial^2 g_{jk}^+}{ \partial z^2} &=-\frac{\hbar^2 k^2}{2 m}g_{jk}^++\bigg[U+\frac{\tilde{U}_j}{1+\rho_j^2}(B_j\rho_j^2+N_j)+\frac{\tilde{U}_{\bar{j}}}{1+\rho_j^2}(C_j\rho_j^2+M_j)\bigg]g_{jk}^+\\
&-\frac{{\rho}_{j}}{1+{\rho}_{j}^2}\bigg[(B_j-N_j)\tilde{U}_j+(C_j-M_j)\tilde{U}_{\bar{j}}\bigg]g_{jk}^-+\frac{Q_j\rho_{\bar{j}}}{\sqrt{\rho_j\rho_{\bar{j}}}(1+\rho_{\bar{j}}^2)}g_{\bar{j}k}^++\frac{Q_j\rho_{\bar{j}}\sqrt{\rho_{\bar{j}}}}{\sqrt{\rho_j}(1+\rho_{\bar{j}}^2)}g_{\bar{j}k}^-\nonumber,
\end{align}
\begin{align}
\frac{\hbar^2}{2 m} \frac{\partial^2 g_{jk}^-}{\partial z^2}&=\frac{\hbar^2 \beta^2}{2 m}g_{jk}^-+\bigg[U+\frac{\tilde{U}_j}{1+\rho_j^2}(B_j+\rho_j^2N_j)+\frac{\tilde{U}_{\bar{j}}}{1+\rho_j^2}(C_j+\rho_j^2M_j)\bigg]g_{jk}^-\\
&-\frac{{\rho}_{j}}{1+{\rho}_{j}^2}\bigg[(B_j-N_j)\tilde{U}_j+(C_j-M_j)\tilde{U}_{\bar{j}}\bigg]g_{jk}^++\frac{Q_j\sqrt{\rho_j}\rho_{\bar{j}}}{\sqrt{\rho_{\bar{j}}}(1+\rho_{\bar{j}}^2)}g_{\bar{j}k}^++\frac{Q_j\rho_{\bar{j}}\sqrt{\rho_{\bar{j}}\rho_{j}}}{(1+\rho_{\bar{j}}^2)}g_{\bar{j}k}^-\nonumber.
\end{align}
\end{widetext}
For $U=0$, equations for $g_{j k}^{\pm}$ are decoupled.  
Therefore, the resulting set of differential equations admit a trivial solution consisting of oscillating plane-wave BQP modes.

From now on we consider a symmetric Bose-Bose mixture where $n_1=n_2=n$, and $g_1=g_2=g$,
which is commonly used in the literature. Therefore, the chemical potential reduces to
	\begin{equation}
 \mu_{\pm}=\delta g_{\pm} n+ \frac{1}{2} \alpha n^{1/2}-\frac{3}{2}\alpha^T n^{-3/2},
	\end{equation}
where $\delta g_{\pm}=g(1\pm g_{12}/g)$.
The corresponding healing length is then given as:
	\begin{equation} \label{HLeng}
\xi_{\pm}=\frac{\hbar^2} {\sqrt{4m\mu_{\pm}}}= \frac{\xi}{ \sqrt{\mu_{\pm}/gn}}=\frac{\hbar} {m c_{s\pm}},
	\end{equation}
where $\xi=\hbar^2/\sqrt{4mgn}$ is the standard healing length of a single BEC.

\begin{figure}
\includegraphics  [scale=0.45] {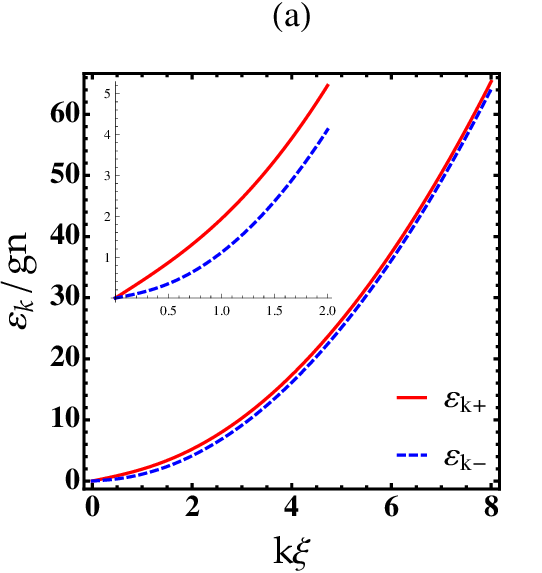} 
\includegraphics  [scale=0.45] {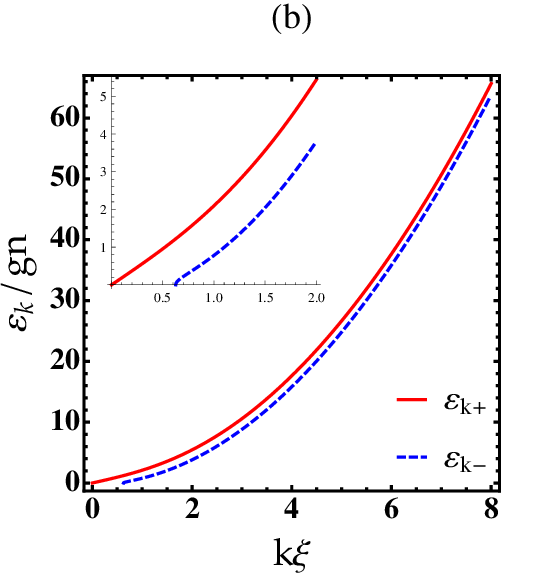} 
\includegraphics  [scale=0.45] {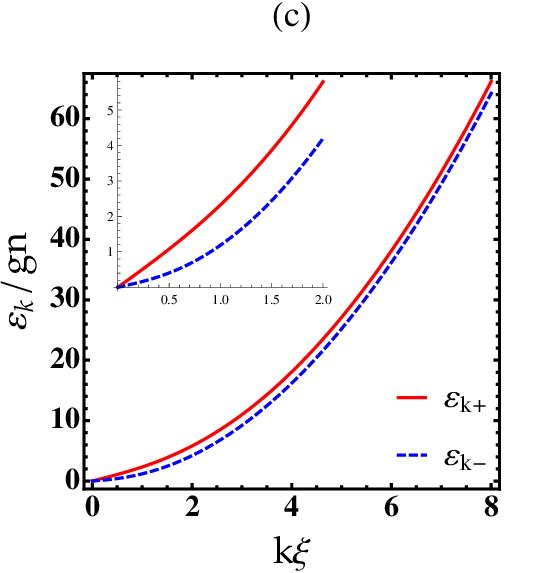} 
\includegraphics  [scale=0.45] {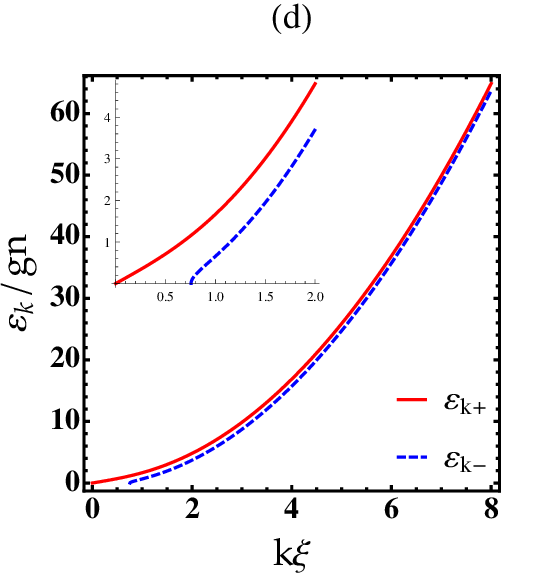} 
\caption{ Bogoliubov spectra, $\varepsilon_{k \pm}$, of a homogeneous Bose-Bose mixture for different values of $\nu$ and $\nu^T$.
(a) ${g_{12}/g}=0.6$,  $\nu=0.2$, and $\nu^T=0.2$.
(b) ${g_{12}/g}=1.5$,  $\nu=0.2$, and $\nu^T=0.2$.
(c) ${g_{12}/g}=1.5$,  $\nu=0.8$, and $\nu^T=0.2$.
(d) ${g_{12}/g}=0.6$,  $\nu=0.2$, and $\nu^T=0.8$.
Here the strength of quantum and thermal fluctuations can be adjusted by means of Feshbach resonance.}
\label{Bogspec}
\end{figure}

For $g_{12}>  g+ \alpha n^{1/2}/2-3\alpha^T n^{-3/2}/2 $, the sound velocity $c_{s-}$ becomes purely imaginary providing an unstable Bogoliubov spectrum. 
This clearly indicates that the presence of the LHY quantum and thermal corrections may modify the stability condition of Bose mixtures.
The Bogoliubov  spectra are plotted in Fig.~\ref{Bogspec}.
For $g_{12}/g <1$ and small $\nu$ and $\nu^T$, both branches, $\varepsilon_{k \pm}$, show the standard Bogoliubov excitations spectrum where
they are phonon-like in the limit $k \rightarrow 0$ (see Fig.~\ref{Bogspec}. a). The same situation holds for $g_{12}/g >1$ but for large quantum fluctuations, $\nu$ 
(see Fig.~\ref{Bogspec}. c). 
However,  for $g_{12}/g >1$, small  $\nu$ and $\nu^T$ and even for $g_{12}/g <1$ and large thermal fluctuations, $\nu_T$, 
the lower branch,  $\varepsilon_{k -}$, becomes imaginary and thus unstable in the low momenta regime (see Fig.~\ref{Bogspec}. b and d). 
Consequently, quantum and thermal fluctuations may stabilize/destabilize weakly repulsive interacting Bose mixtures in a random environment giving rise to
affect the miscibility condition and the localization process of BQPs.

\begin{figure}
\includegraphics  [scale=.6] {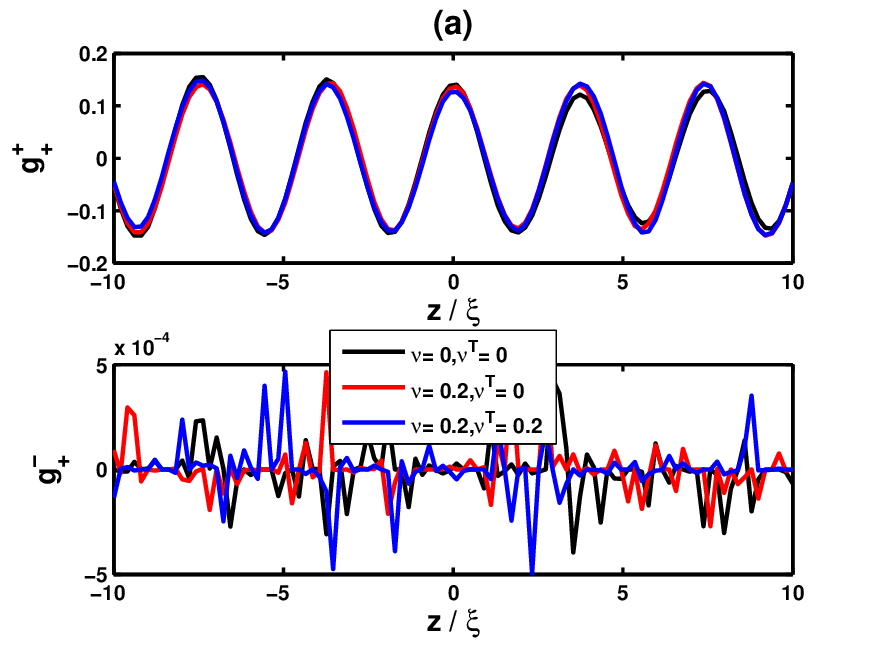} 
\includegraphics  [scale=.6] {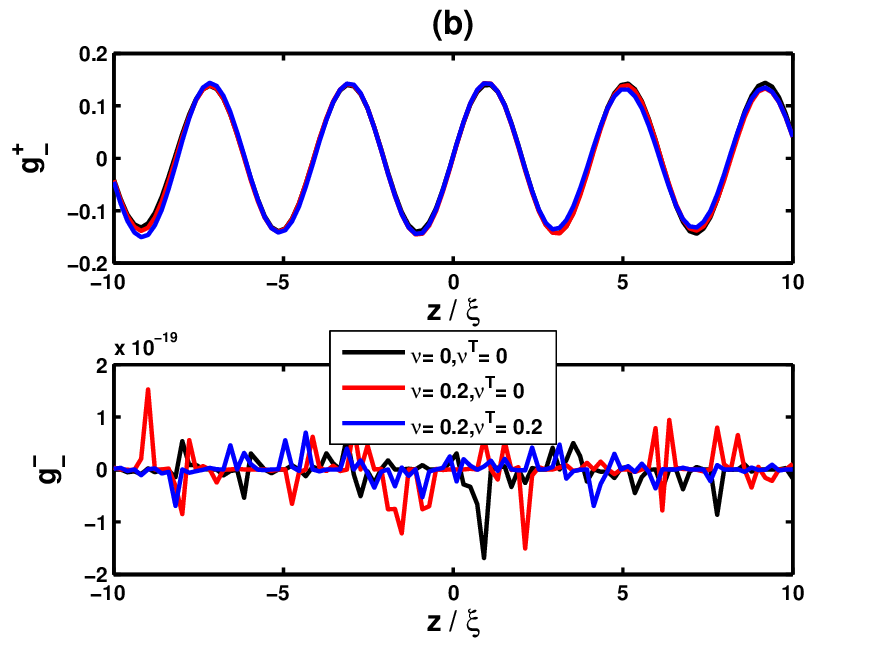} 
\caption{BQP modes in the $g_{\pm}$ basis at energy $\varepsilon/ng=1.1$ with $g_{12}/g=1.5$, $ \sigma/\xi=0.2$, and $U_{R}/ng=0.1$.}
\label{BogM}
\end{figure}

In the case of a weak disorder, equations for $g_{k\pm}^{\pm}$ are coupled  by a term of the order of $2\rho_{jk} \tilde U_{\pm}/(1+\rho_{jk}) \leq U$ enabling us to neglect
$g_{k \pm}^{-}$ to first-order in disorder (i.e. $|g_{k\pm}^{-}| \ll |g_{k \pm}^{+}|$). 
This assumption is confirmed by our numerical results as shown in Fig.~\ref{BogM}.
We see also from the same figure that the functions $g_{k \pm}^{-}$ vary rapidly and on a length scale smaller than the healing length regardless of the LHY strength.

Therefore, the transport properties are fully determined by the functions $g_{k \pm}^{+}$:
\begin{equation}  \label{EShro}
-\frac{\hbar^2}{2m} \frac{\partial^2g_{k\pm}^+}{\partial z^2} +{\cal U}_{k\pm}(z)g_{k \pm}^+  \simeq E_k g_{k \pm}^+,
\end{equation}
where 
\begin{equation} 
{\cal U}_{k\pm}(z)=U (z)-\frac{\tilde{U}_{\pm}(z)}{1+\rho_{\pm}^2}\bigg(3+\rho_{\pm}^2-\frac{\alpha n^{1/2}}{2\mu_{\pm}}+\frac{15\alpha^Tn^{-3/2}}{2\mu_{\pm}}\bigg).
\end{equation}
Here, for convenience, we chose the index $\pm$ instead of $j$ for the quantities: $\rho$, $\tilde{U}$, ${\cal U}_{k}$, and $g_k^+$.\\
Importantly Eq.~(\ref{EShro}) is formally equivalent to a Schr\"odinger equation
for a bare particle of energy $E_k$ in an effective potential ${\cal U}_{\pm}(z)$. 
This latter which depends on the BQPs energy $\varepsilon_{k\pm}$ via  $\rho_{k \pm}$ and valid for any weak external potential, differs from both the bare potential $U (z)$ 
and the smoothed potential $\tilde U_{\pm} (z)$.
Evidently, the original system of disordered interacting particles can be described by Eq.~(\ref{EShro}) for noninteracting quasiparticles in disordered media.

\section{Localization of excitations}  \label{ALE} 

\begin{figure}
\includegraphics  [scale=0.35] {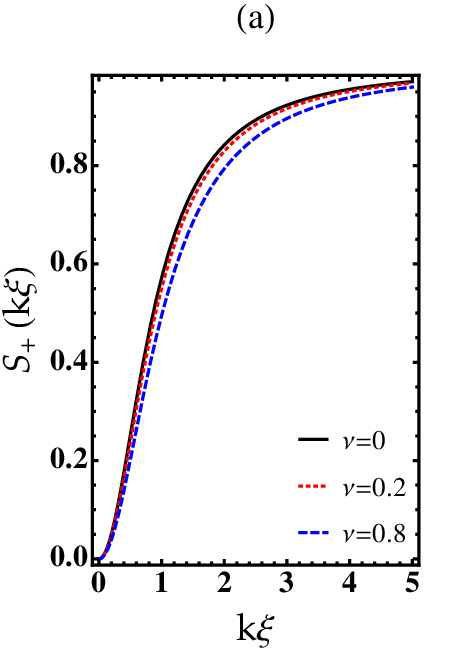}
\includegraphics  [scale=0.35] {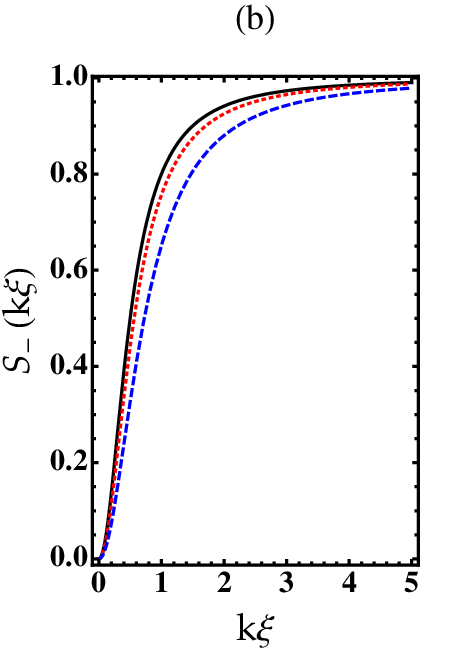} 
\includegraphics  [scale=0.35] {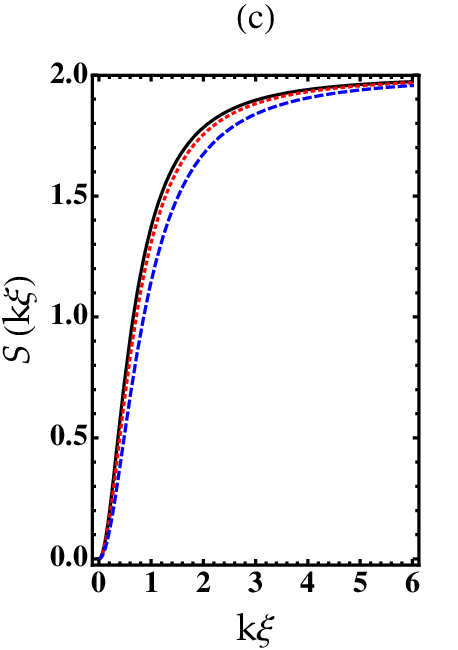} 
\includegraphics  [scale=0.35] {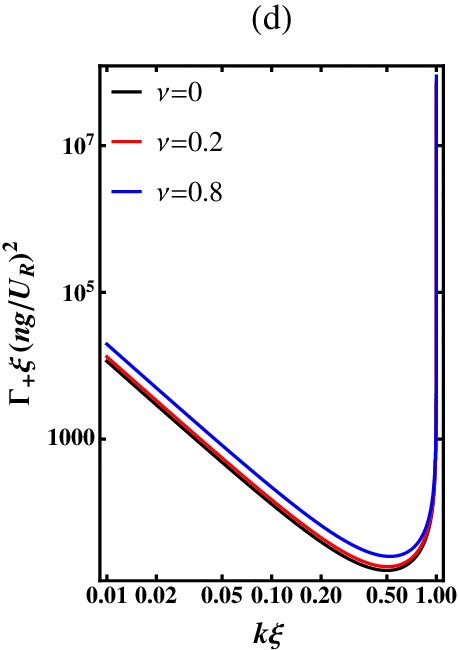}
\includegraphics  [scale=0.35] {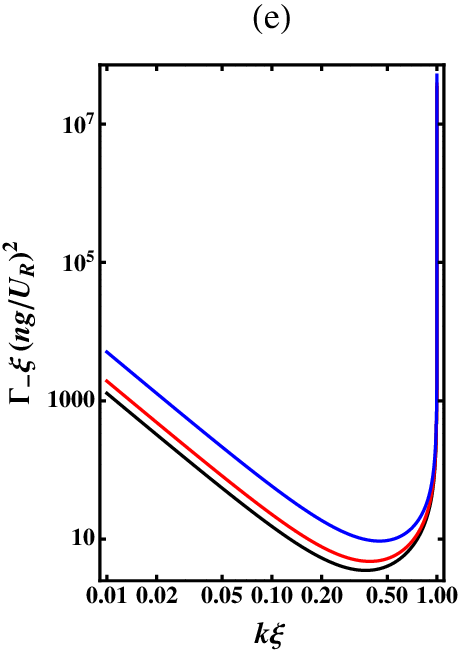} 
\includegraphics  [scale=0.35] {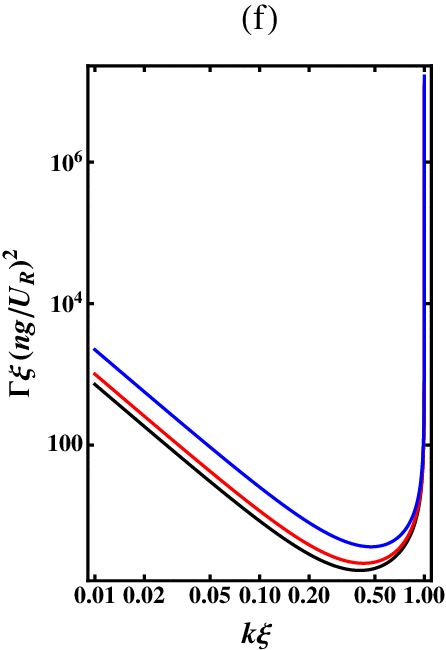}
\caption{Screening functions (a) $S_+$, (b) $S_-$, and  (c) $S=\sum\limits_{\pm} S_{\pm}$ for different values of $\nu$ at zero temperature (i.e. $\nu^T=0$).
Lyapunov exponent (a) $\Gamma_+$, (b) $\Gamma_-$, and  (c) $\Gamma=\sum\limits_{\pm} \Gamma_{\pm}$ for different values of $\nu$ at zero temperature.
Parameters are: $g_{12}/g=0.6$ and  $\sigma/\xi=1$.}
\label{LET0}
\end{figure}

\begin{figure}
\includegraphics  [scale=0.35] {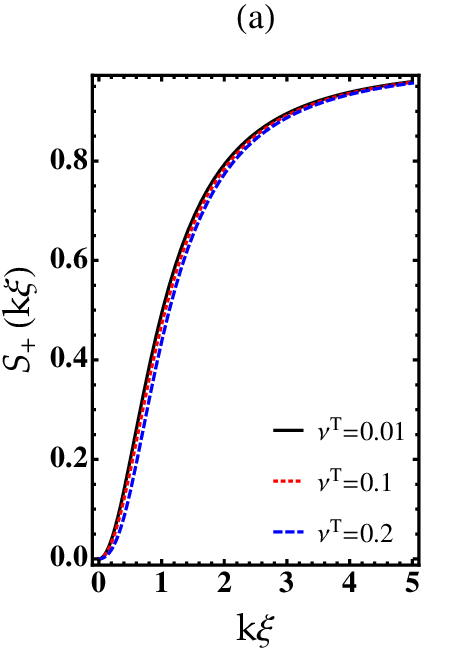}
\includegraphics  [scale=0.35] {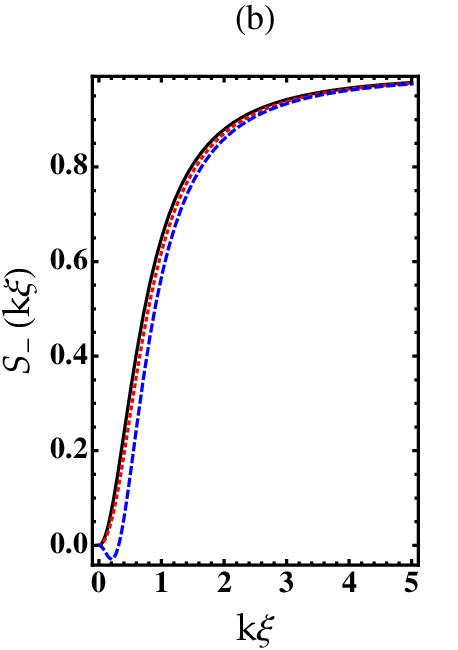} 
\includegraphics  [scale=0.35] {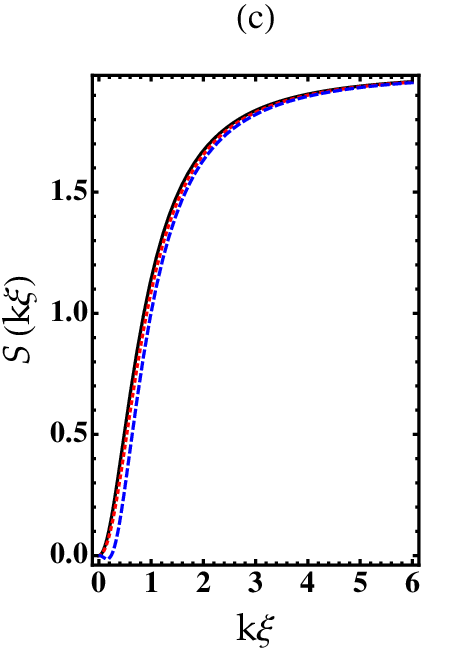} 
\includegraphics  [scale=0.35] {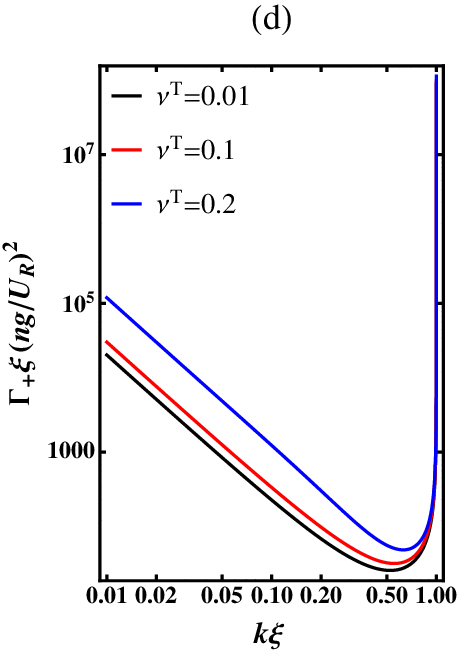}
\includegraphics  [scale=0.35] {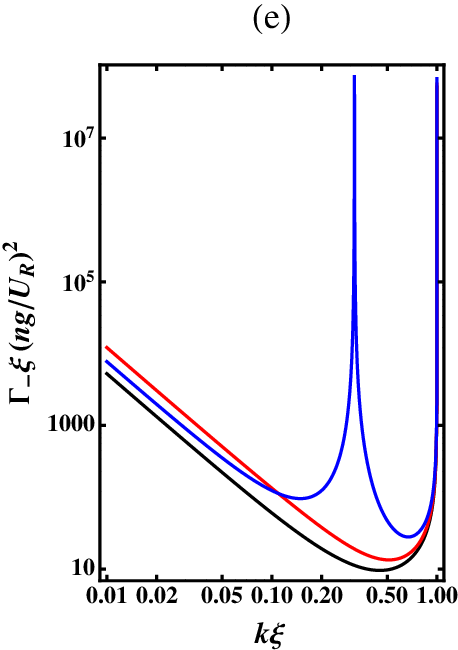} 
\includegraphics  [scale=0.35] {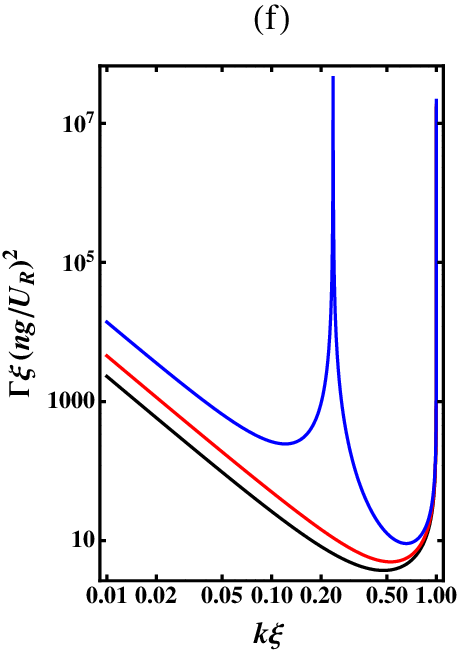} 
\caption{The same as Fig.~\ref{LET0} but in the presence of thermal fluctuations with $\nu=0.8$.}
\label{LET}
\end{figure}

\begin{figure}
\includegraphics  [scale=0.3] {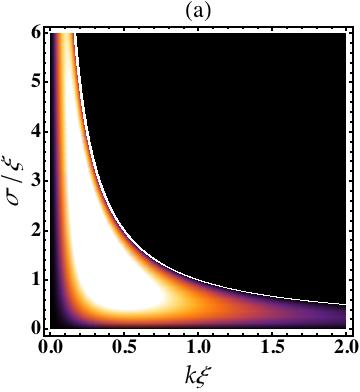}
\includegraphics  [scale=0.3] {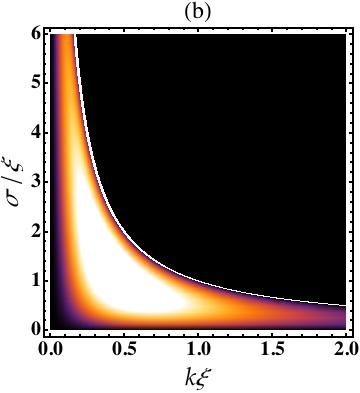} 
\includegraphics  [scale=0.3] {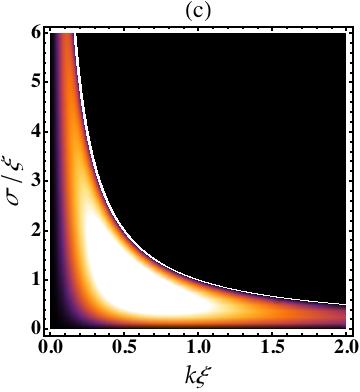} 
\includegraphics  [scale=0.3] {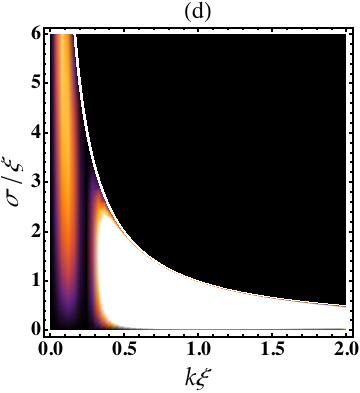}
\caption{ Density plot of the total Lyapunov exponent, $\Gamma=\sum\limits_{\pm}\Gamma_{\pm}$, of BQPs in a 1D speckle potential for sevral values of $\nu$ and $\nu^T$ 
with $g_{12}/g=0.6$.
(a) $\nu=\nu^T=0$. (b) $\nu=0.8$ and $\nu^T=0$. (c) $\nu=0.8$ and $\nu^T=0.1$. (d) $\nu=0.8$ and $\nu^T=0.2$.
White dashed lines represent an effective mobility edge due to the cutoff in the power spectrum of the disorder.}
\label{lupy}
\end{figure}

In 1D systems, the solution of Eq.~(\ref{EShro}) in the frame of the on-shell approximation \cite{Cord} yields for the Lyapunov exponent 
corresponding to the density, $\Gamma_{+}$, and spin, $\Gamma_{-}$, parts of BQPs
	\begin{align} \label{Lup1D}
\Gamma_{\pm} = \frac{1}{\ell_{\pm}} \sim \frac{\pi}{8} \left(\frac{U_R}{ng}\right)^2 \frac{\sigma_R}{k^3\xi^4}S_{\pm}^2(k\xi)\hat{c}(2k\sigma), 
	\end{align}
where $\ell_{\pm}$ is the localization length in each component, and 
	\begin{align} \label{UF}
		S_{\pm}(k\xi)&=\frac{2k^2\xi^2}{2k^2\xi^2+\delta g_{\pm}/g+\nu/2-3\nu^T/2} \\
&\times \bigg[1+\frac{\nu/2-15\nu^T/2}{2(4k^2\xi^2+\delta g_{\pm}/g+\nu/2-3\nu^T/2)}\bigg], \nonumber
	\end{align}
are the universal functions $S_{\pm}$ which define the screening terms \cite{Laurent2}. They encode both the interaction and the LHY effects.
In the phonon regime where $k\xi \ll 1$, one has 
$$S_{\pm}^2(k \xi) \sim \frac{\left(2 \delta g_{\pm}/g+3 \nu/2 -21 \nu^T/2 \right)^2}{ \left(\delta g_{\pm}/g+\nu/2 -3 \nu^T\right/2)^4} k^4 \xi ^4,$$ 
indicating that the disorder is crucially screened.
In the limit of high energy $k\xi \gg 1$, $S_{\pm}^2 (k \xi) \sim 1$ reproducing the free-particle case. 
Importantly, the dimensionless Lyapunov exponents $\Gamma_{\pm}$ (\ref{Lup1D}) rely on four parameters: $k\xi$, $\sigma/\xi$, $\nu$, and $\nu^T$ unlike the
the standard form of $\Gamma$  which depends only on $k\xi$, $\sigma/\xi$ \cite{Laurent2}.
For $g_{12}=0$, the localization length reduces to that of a single  BEC with the LHY corrections (see Appendix B).

For the sake of concreteness, we will consider optical speckle potentials, which are often used in ultracold atoms experiments
\cite{Lye, Billy, Roat}. The autocorrelation functions of such speckles are given in the Fourier space  as \cite{Cord} 
\begin{equation}\label{ddp}
\hat c_2({k}\sigma)= \sqrt{\frac{\pi}{2}} \left(1-\frac{k \sigma}{2} \right) \Theta\left(1-\frac{k\sigma}{2}\right),
\end{equation}
 where $\Theta$ is the Heaviside function. 
The Heaviside distribution shows that the disorder potential is smooth on length scales smaller than the healing length, while it is uncorrelated elsewhere.
For $k \sigma \ll 1$, the autocorrelation function (\ref{ddp}) reduces to that of the white-noise disorder, $\hat c_2({q})=\hat c_2(0)$.  
Thus, $\Gamma_{\pm}$ can be calculated analytically in the full spectrum of interaction regimes.

The behavior of the functions $S_{\pm}$ and their influence on the localization length $\Gamma_{\pm} = 1/\ell_{\pm}$ in each component  at zero temperature 
for different values of $\nu$ are shown in Fig.~\ref{LET0}.
We see from Fig.~\ref{LET0} (a)-(c) that important changes occur in the screening functions, $S_{\pm}$, when
the elementary excitations are transformed from the phonon regime ($k\xi \ll 1$) to free-particle domain  ($k\xi \gg 1$)  
regardless of the strength of the LHY quantum correction, $\nu$. 
Similarly  to the single BEC  case \cite{Laurent2},  the functions $S_{\pm}$ substantially decay in the phonon regime due to the high screening of the disorder potential by the density background. 
In the crossover region which is characterized by competition of the bare kinetic energy, mean-field and beyond mean-field (LHY) corrections, 
the function $S_-$ corresponding to the lower branch of excitations is critically sensitive to the LHY quantum term, $\nu$, 
compared to the screening function of the upper branch, $S_+$.
This can be explained by the considerable LHY quantum effect which has a tendency to decrease $\varepsilon_{k-}$, thus boosting the contribution of the spin excitations 
to the screening function $S_-$ giving rise to reduce the localization in such a component.

Figures \ref{LET0} (d)-(f) show that the Lyapunov exponents, $\Gamma_{\pm}$, diverge at low energy due to screening, and diverge as well at high energy (free-particle regime). 
Thus,  they exhibit a local minimum around $k_{\pm}^{ \text{min}} \xi= \sqrt{2\delta g_{\pm}/g-3\nu/2-2}/(2\sqrt{2})$ at intermediate energy.  
The position and the depth of the minimum in $\Gamma_{\pm}$ and $\Gamma$ do not depend only on the disorder correlation function but also depend on the LHY quantum corrections.
Rising the strength of the LHY term  leads to up shift  all the localization lengths in the whole range of wavenumber-$k$.
As expected, the Lyapunov exponent corresponding to the lower branch $\Gamma_{-}$ is more affected by the LHY quantum fluctuations.
Therefore, the localization of the density excitations is strengthened by the screening effects compared to the spin excitations component.
One can infer that the AL of BQPs in one component does not necessarily trigger the localization in the other component.

At finite temperatures, the situation is quite different. 
Figure \ref{LET} shows that the thermal fluctuations may substantially affect the standard localization picture in particular in the spin excitations component.
Remarkably, the screening function $S_-$ develops a dip at very low momentum owing to the thermal fluctuation effects which tend to acutely modify the spin part of the Bogoliubov excitations.
Its corresponding localization length $\Gamma_-$ exhibits two minima revealing the ocurrence of two localization maxima (see also Fig.~\ref{lupy} (d)). 
This can be attribute to (i) the strong dependence of $\Gamma_-$ on the momentum of the spectrum which itself follows the thermal fluctuation modulations,
(ii) the significant screening in the phonon regime.

To gain deep insights into localization properties, we analyze the behavior of the total Lyapunov exponent (\ref{Lup1D}) including the function (\ref{ddp}). 
The results are presented in Fig.~\ref{lupy} in plan $(k\xi, \sigma/\xi)$.
We see that $\Gamma=\sum_{\pm}\Gamma_{\pm}$ vanish in the Born approximation giving rise to an effective mobility edge, 
due to a high-momentum cut-off of the speckle power spectrum (specific correlation properties of speckle potentials) \cite{Laurent2}. 
Inceasing the thermal fluctuations leads to induce two and eventually multiple localization maxima even with standard speckle correlations. 
Strong enough thermal fluctuations may dominate the disorder prohibiting the localization notably in the lower excitations branch resulting in two or more extended (delocalized) states.
Note that such localization maxima have been also observed for free particles \cite{Plod,Piraud} using for example crossed Gaussian beams.

\section{Anderson localization of BQP of quantum liquids}  \label{ALQL} 

In this Section we apply the above formalism in order to check the existence of AL of BQPs in 1D mixture quantum droplets.
For simplicity, we consider here a sufficiently large droplet, where the edge effect can be safely neglected. Consequently, the condensate wavefunction becomes a constant.
Assuming that the equilibrium densities of both components are identical, which renders the analysis tractable. 
Therefore, the GGPE (\ref{GPE}) takes the form \cite{Petrov1,Boudj5}:
	\begin{equation} \label{DGPE}
		i\hbar\frac{\partial\phi}{\partial t}=\bigg(-\frac{\hbar^2}{2m} \frac{\partial^2 }{\partial z^2}+U+\delta g|\phi|^2 -\frac{\sqrt{2m}}{\pi \hbar}g^{3/2}|\phi|\bigg)\phi,
	\end{equation} 
where  the coupling constant $g>0$ is relevant for inducing a hard mode while $\delta g_{+}$ is responsible for appearance of a soft  mode.
The condition $\delta g/g=\delta g_{+}/g > 0 $ is necessary for obtaining an energy minimum in the  ground-state energy \cite{Petrov1}. 

Using the above density-phase picture,  the hydrodynamic equations corresponding to the GGPE (\ref{DGPE}) turn out to be given as:
	\begin{equation}
		\bigg(-\frac{\hbar^2}{2m} \frac{\partial^2}{\partial z^2}  + U -\mu+\delta g n-\frac{\sqrt{2m}}{\pi \hbar}g^{3/2} n^{1/2}\bigg)\sqrt{n}=0,
	\end{equation}
	\begin{equation}
		\hbar\frac{\partial}{\partial t}\frac{\delta n_c}{\sqrt{n}} =\bigg(-\frac{\hbar^2}{2m} \frac{\partial^2}{\partial z^2}  + U -\mu+\delta g n-\frac{\sqrt{2m}}{\pi \hbar}g^{3/2} n^{1/2}\bigg)2\sqrt{n}\theta,
	\end{equation}
and
	\begin{equation}
		-2\sqrt{n}\hbar\frac{\partial \theta}{\partial t}=\bigg(-\frac{\hbar^2}{2m} \frac{\partial^2}{\partial z^2}  + U -\mu+3 \delta g n-2\frac{\sqrt{2m}}{\pi \hbar}g^{3/2} n^{1/2}\bigg)\frac{\delta n_c}{\sqrt{n}}.
	\end{equation}
The BdGE can be found by expanding the phase and density operators via (\ref{B:td0}). This yields
	\begin{equation}
		\varepsilon_k f_k^-=\bigg(-\frac{\hbar^2}{2m} \frac{\partial^2}{\partial z^2}  + U -\mu+\delta g n-\frac{\sqrt{2m}}{\pi \hbar}g^{3/2} n^{1/2}\bigg)f_k^+,
	\end{equation}
	\begin{equation}
		\varepsilon_k f_k^+=\bigg(-\frac{\hbar^2}{2m} \frac{\partial^2}{\partial z^2}  + U -\mu+3\delta gn-2\frac{\sqrt{2m}}{\pi \hbar}g^{3/2} n^{1/2} \bigg)f_k^-. 
	\end{equation}
For sufficiently weak disorder, we can use  the above perturbation expansion around the obtained BdGE
which govern the quasi-particle excitations of  a large droplet. Then,  we introduce the same transformation that maps the
many-body Bogoliubov equations onto an effective Schr\"odinger equation which describes the scattering of BQPs in weak disorder potentials.
After a straightforward calculation, we get the following closed equation valid to first-order in $U$
	\begin{equation}
		-\frac{\hbar^2}{2m} \frac{\partial^2g_{k}^+}{\partial z^2} + {\cal U}_{k} (z) g_{k}^+\simeq \frac{\hbar^2 k^2}{2m}g_{k}^+,
	\end{equation}
	where
	\begin{equation}
		{\cal U}_{k} (z)=U(z)-\frac{\tilde{U}(z)}{1+\rho^2}\bigg(3+\rho^2+\frac{\sqrt{2m}}{2\mu\pi \hbar}g^{3/2}n^{1/2}\bigg),
	\end{equation}
is an effective potential relies on the BQPs energy and has a vanishing average.
In Fourier space, it can be written as: ${\cal U}{(2k)}=S(k\xi) U(2k)$,
	where
	\begin{equation} \label{screeDrop}
	S(k\xi)=\frac{2k^2\xi^2}{1+2k^2\xi^2-\beta/2}\bigg[1-\frac{\beta/2}{2(1+4k^2\xi^2-\beta/2)}\bigg],
	\end{equation}
where $\beta=\sqrt{2m} g^{3/2} n^{-1/2}/ (\pi \hbar \delta g)= N_{\text{cr}}/(n \xi)$ with 
$N_{\text{cr}}= (g/\delta g)^{3/2} (\sqrt{2}/\pi)$ being the critical atom number separating  BEC and droplet phases, and $\xi= \hbar/\sqrt{ nm \delta g}$.
In an infinitely extended uniform liquid (large $N_{\text{cr}}$),  $\mu= -4mg^3/(9 \pi^2\hbar^2 \delta g)$. 
In such a case, the droplet features an exponentially weak dependence on the number of particles \cite{Astra,Boudj12}.
The Lyapunov exponent of the BQPs of a droplet in a 1D weak disordered potential is then given by :
	\begin{equation} \label{LypDrop} 
		\Gamma_k \sim\frac{\pi}{8}\bigg(\frac{U_R}{\mu}\bigg)^2 \frac{\sigma_R}{k^2 \xi^4} [S(k\xi)]^2\hat{c_2}	(2k\sigma_R).
	\end{equation}
As expected in the free-particle regime, $k\xi \gg1$, $S(k\xi) \sim 1$ and hence,  $\Gamma_k \sim 1/k^2$ meaning that the AL of BQPs can be described exactly by
the  Schr\"odinger equation. 
Oppositely, for $k\xi \ll 1$, we have $S(k\xi) \sim 2 (4-3\beta) k^2\xi^2/(2-\beta)^2$. Therefore, the Lyapunov exponent  (\ref{LypDrop}) reduces to 
$\Gamma_k \sim \pi \big(U_R/\mu\big)^2 [(4-3\beta)/(2-\beta)^2]^2 \sigma_R \hat{c_2} (2k\sigma_R) k^2$ which differs from the standard 
localization of  phonons in 1D BEC \cite{Laurent2} by a factor $(4-3\beta)/(2-\beta)^2$.
For $\beta \geq 2$, the screening becomes infinite and thus AL of BQPs does not take place due to the disappearance of the droplet excitations (full delocalization).

\begin{figure}
\center
\includegraphics  [scale=0.8] {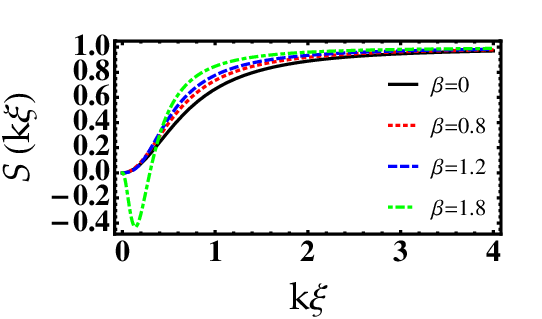}
\includegraphics  [scale=0.25] {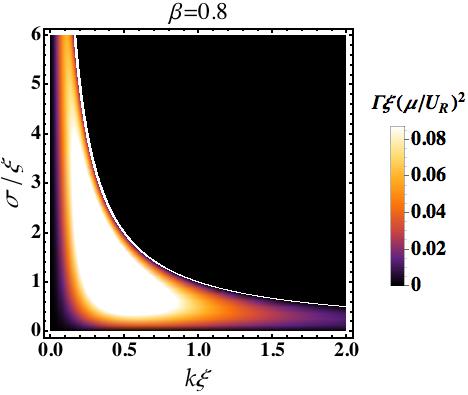} 
\includegraphics  [scale=0.25] {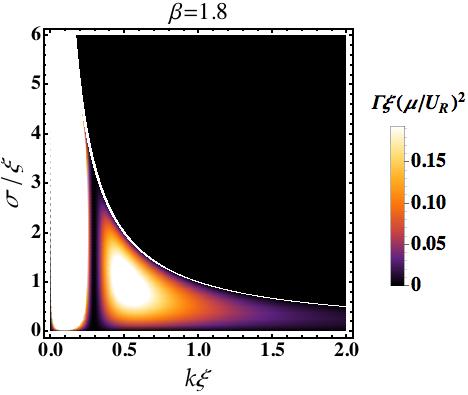} 
\caption{ Screening function from Eq.~(\ref{screeDrop}) for different values of $\beta$. Dashed line corresponds to the screening function of an ordinary BEC for comparison.
Density plot of the Lyapunov exponent of droplet BQPs in a 1D speckle potential for different values of $\beta$.
Here the white dashed lines represent an effective mobility edge due to the cutoff in the power spectrum of the disorder.}
\label{locdrp}
\end{figure}

The function $	S(k\xi)$ (\ref{screeDrop}) and the  Lyapunov exponent (\ref{LypDrop}) are plotted  in Fig.~\ref{locdrp} 
for a 1D speckle potential with reduced autocorrelation function given by Eq.~(\ref{ddp}).
We observe that the Lyapunov exponent goes to infinity at low momenta and hence the BQPs become less localized when the droplet reaches its flat-top regime. 
Only small fraction of BQPs is localized at intermediate momenta.
This is most likely due to the fact that in the self-bound droplet, the energy of collective excitations is inherently bounded from above by the particle-emission threshold 
meaning that it cannot survive in an excited state, giving rise to reduce the AL of BQPs.
However, AL of BQPs remains robust in the small droplet regime (i.e. for small $\beta$, or equivalently small $N_{\text{cr}}$).

\section{Conclusions} \label{Conc}

In this paper, we theoretically studied AL of BQPs in a weakly interacting Bose mixture with the LHY quantum and thermal fluctuations placed into 1D  weak and correlated random potentials.
The theoretical approach employed here is based on a weak-disorder perturbative expansion.
Our numerical and analytical results revealed that the competition between disorder, and the LHY corrections  leads to modify the density profiles, 
the Bogoliubov modes, and the localization length in each species.  
For 1D speckle potentials, we found that the BQPs undergo genuine AL in the density excitations component while they are less localized in the spin excitations component.
This latter favors several localization maxima in the presence of strong thermal fluctuations even without any tailored exotic disorder correlations.
In the case of 1D self-bound liquids, we showed that the BQPs exhibit a very weak AL localization in the plateau region. 
Furthermore, we emphasized that when the LHY terms exceed a certain critical value, the BQPs become fully delocalized. 
One should stress that the obtained results can be readily extended to higher dimensions.
A natural extension of our work is to treat AL of BQPs for droplets loaded in optical lattices (discrete droplets) with strong disorder.

\section*{Acknowledgements}

We thank Laurent Sanchez-Palencia and Axel Pelster for stimulating discussions.

\newpage 
\appendix
\include{appendixA} \label{appendixprocedure}
\renewcommand{\theequation}{A\arabic{equation}}
\renewcommand{\thefigure}{A\arabic{figure}}

\section{Equation of state and compressibility}  \label{A}
In this Appendix we calculate corrections induced by the disorder to the EoS and to the compressibility.
Let us start by writing perturbative functions $\phi^{(i)}$ in momentum space. 

	\begin{equation} \label{eq1}
	\phi^{(0)}=1,
	\end{equation}
	\begin{equation}\label{eq2}
	\phi^{(1)}(k)=-\frac{1}{2}G_{\xi}*\frac{U (k)}{\mu_0},
	\end{equation}
and 
	\begin{align}\label{eq3}
	\phi^{(2)}(k)&=-\frac{1}{2}\int \frac{d k'} {2\pi} G_{\xi_j}*\bigg[\frac{U( k')}{\mu_0}+\frac{3\delta gn\phi^{(1)}( k')}{\mu_0}\\
&+\frac{\alpha n^{(1/2)}\phi^{(1)}( k')}{\mu_0}\bigg]\phi^{(1)} ( k-k'), \nonumber
	\end{align}
	where $\mu_0=\delta g n+\frac{1}{2}\alpha n^{1/2}$.
The total density of the fluid is defined as:
		\begin{equation} \label{eq4}
n= \langle \phi^2(z) \rangle=n_{0}\bigg(\phi^{(0)}+  \langle\phi^{(1)2}(z) \rangle+2\phi^{(0)} \langle \phi^{(2)} (z) \rangle\bigg).
	\end{equation}
The EoS can be calculated by substituting Eqs.~(\ref{eq1})-(\ref{eq3}) into Eq.~(\ref{eq4}) and solving the equation $\langle \phi^2(\mu) \rangle$,
	where $\mu$ represents the bare chemical potential.
This immediately gives:
	\begin{align} \label{eq5}
\mu&=\delta g n+\alpha n^{1/2}\\
&-\int\frac{d k}{ 2\pi}\frac{E_kR(k)}{\bigg[E_k+2\left(\delta g n+\alpha n^{1/2}/2\right)\bigg]^2}. \nonumber
\end{align}
One should stress that for a delta-correlated disorder the chemical potential (\ref{eq5}) needs to be renormalized \cite{Boudj6}.

The inverse compressibility is  defined as $\kappa^{-1} = n^2 \left(\partial \mu/\partial n \right)$, where
	\begin{align}\label{eq6}
\frac{\partial \mu}{ \partial n}&=\delta g+\alpha n^{-1/2}/2\\
&+\int\frac{d k}{ 2\pi}\frac{4 E_k (\delta g+\alpha n^{-1/2}/4) R(k)}{\bigg[E_k+2\left(\delta g n+\alpha n^{1/2}/2\right)\bigg]^3}. \nonumber
\end{align}
This equation is important since it allows us to calculate the equilibrium critical density of the quantum liquid. 
It enables us also to distinguish quantum droplets from bright solitons.

For a speckle disorder with correlation function defined in Eq.~(\ref{ddp}), we obtain for the EoS
\begin{equation}
\frac{\mu}{\delta g n}=1+2 \nu- R_0 f(\sigma/\xi, \nu),
\end{equation}
where $R_0= U_R^2/(\delta g n)^2$ and 
	\begin{align}\nonumber
f(\sigma/\xi, \nu)&=\frac{1}{4\sqrt{2\pi}} \bigg[\frac{\xi^2 } {4 \sigma ^2}  \ln \left(\frac{\nu +1} {\nu +4 \sigma ^2/\xi ^2+1}\right)  \\
 &+\frac{(\xi/2 \sigma)   \text{Arctan}\left(\frac{2 \sigma/\xi } {\sqrt{\nu+1} }\right)}{\sqrt{\nu +1} } \bigg], \nonumber
\end{align}
is the disorder function, its behavior is displayed in Fig.~\ref{DisF} (a).

 \setcounter{figure}{0}
\begin{figure}
\includegraphics  [scale=0.45] {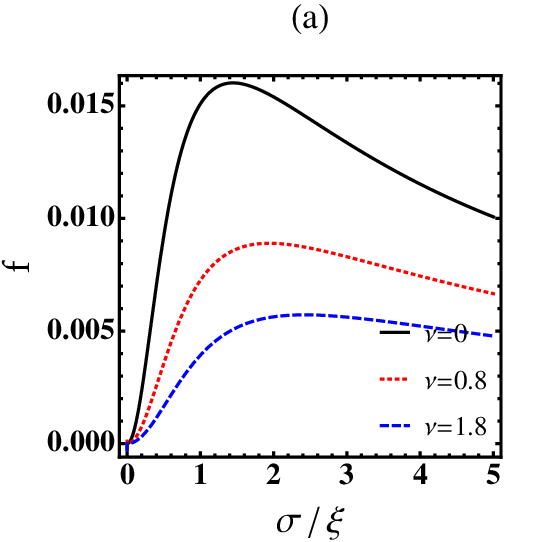}
\includegraphics  [scale=0.45] {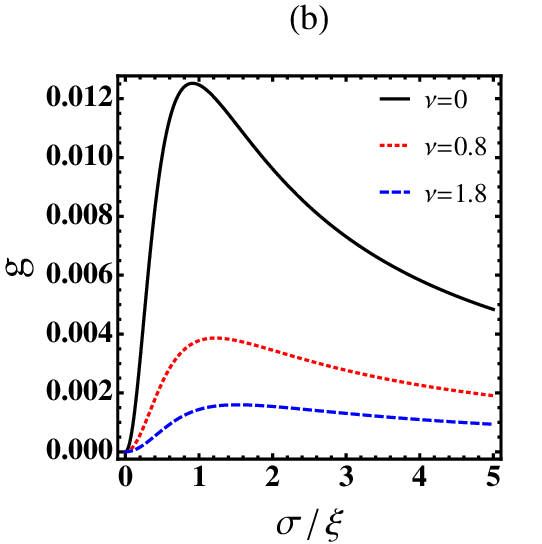} 
\caption{ Disorder function  corresponding to the EoS, $f(\sigma/\xi, \nu)$, and to the compressibility, $g(\sigma/\xi, \nu)$, as a function of  $\sigma/\xi$ for different values of $\nu$.}
\label{DisF}
\end{figure}

The inverse compressibility (\ref{eq6}) turns out to be given as:
	\begin{align}
\frac{\partial \mu}{ \partial n}&=\left(\delta g+\alpha n^{-1/2}/2 \right) \left[1+ R_0  g(\sigma/\xi, \nu) \right],
\end{align}
where the disorder function is defined as
\begin{align}\nonumber
 g(\sigma/\xi, \nu)=\frac{ (\xi/2 \sigma) \left(\nu +\frac{4 \sigma ^2}{\xi ^2}+1\right) \text{Arctan}\left(\frac{2 \sigma/\xi  }{\sqrt{\nu +1} }\right)-\sqrt{\nu +1}}
{16 \sqrt{\pi} (\nu +1)^{3/2} \left(\nu +\frac{4 \sigma ^2}{\xi ^2}+1\right)},
\end{align}
and its behavior is shown in Fig.~\ref{DisF} (b).

Figure \ref{DisF} depicts that the main disorder  contribution to the EoS and to the compressibility comes from the region  $\sigma \sim \xi$.
We see also that the presence of the LHY quantum fluctuations lead to reduce the disorder effects in such quantities.
 
\include{appendixB} 
 
\renewcommand{\theequation}{B\arabic{equation}}
\renewcommand{\thefigure}{B\arabic{figure}}

\section{AL of BQPs in a disordered single  BEC with the LHY corrections}   \label{B}

In this appendix we consider the case of $g_{12}=0$ in which the system is uncoupled.
We look in particular at how the competition between the random potential and LHY zero and finite-temperature corrections affects the AL of BQPs.

Using our FTGGPE and the hydrodynamic model for the BQPs and setting $g_{12}=0$, we find for the Lyapunov exponent
\begin{equation} \label{SLE}
\Gamma\sim \frac{\pi}{8}\bigg(\frac{U_R}{ng}\bigg)^2\frac{\sigma_R}{k^3\xi^4}S^2(k\xi)\hat{c}(2k\sigma_R),
\end{equation}
where  the universal screening function is given by
\begin{align} 
S(k\xi)&=\frac{2k^2\xi^2}{2k^2\xi^2+1+\nu/2-3\nu^T/2} \\
&\times\bigg[1+\frac{\nu/2-15\nu^T/2}{2(4k^2\xi^2+1+\nu/2-3\nu^T/2)}\bigg]. \nonumber
\end{align}

The screening function, $S(k\xi)$, and the  Lyapunov exponent, $\Gamma$, of Eq.~(\ref{SLE}) are plotted in Fig.~\ref{SingLET} for a 1D speckle potential given by Eq.~(\ref{ddp}).
It is clearly visible that the presence of quantum and thermal fluctuations may shift $S(k\xi)$ and the position of the minimum appearing in $\Gamma$.
This may lead to crucially alter the disorder-induced localization of BQPs in a weakly interacting single Bose gas.

 \setcounter{figure}{0}
\begin{figure}
\includegraphics  [scale=0.45] {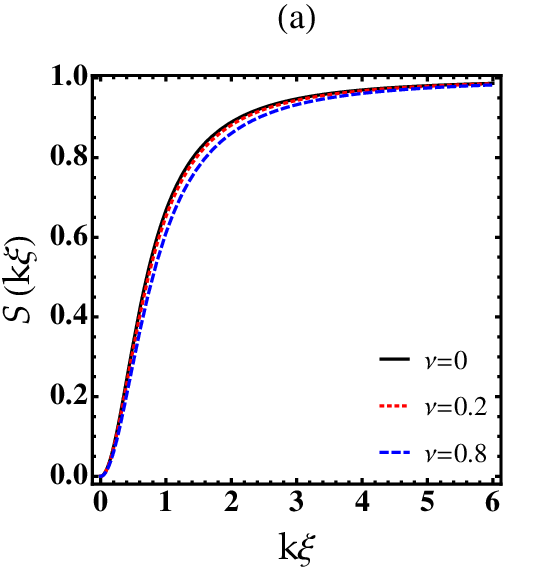}
\includegraphics  [scale=0.45] {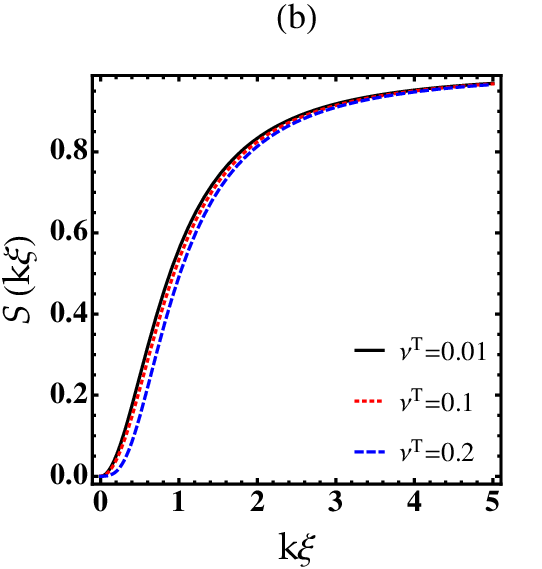} 
\includegraphics  [scale=0.45] {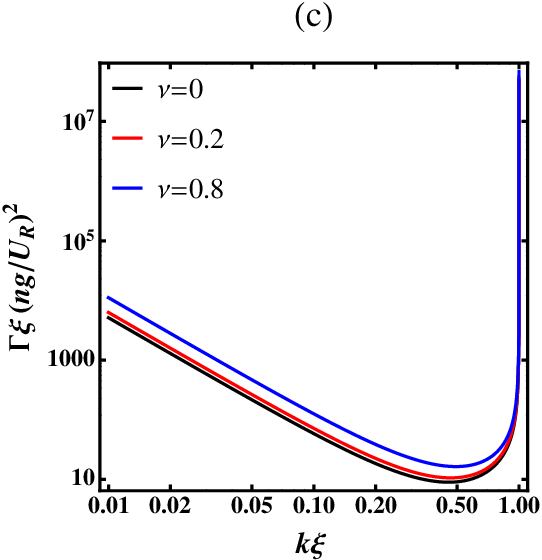}
\includegraphics  [scale=0.45] {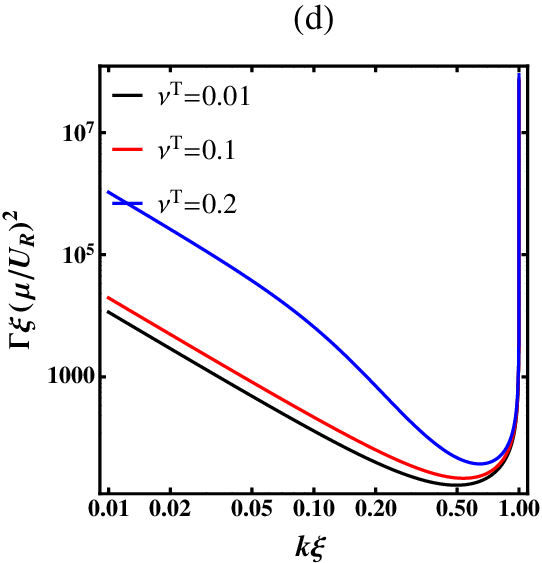} 
\caption{Screening function, $S(k\xi)$, at  zero temperature (i.e. $\nu^T=0$) (a), and  at finite temperatures with $\nu=0.8$ (b).
Lyapunov exponent, $\Gamma \xi (ng/ U_R)^2$, at zero temperature (c), and at finite temperatures with $\nu=0.8$ (d).
Parameters are: $g_{12}/g=0$ and  $\sigma/\xi=1$.}
\label{SingLET}
\end{figure}

\end{document}